# Human and Artificial Intelligence – Promoting Trustworthy and Understandable Collaboration

*Gilbert Drzyzga*[1]

Methods of Artificial Intelligence (AI) enable the personalization of information for individual user experiences in many domains; however, they can also conflict with established design principles, e.g., due to uncertainties regarding the real world. Building trust and understanding can serve as an approach to create a more balanced relationship between humans and AI. Building upon a pilot study, an online survey was conducted to investigate 12 individual aspects related to the topics of explainability and controllability. The results indicate that both topics, despite their different and numerous facets, are generally perceived as important by respondents; simultaneously, however, a wide dispersion of opinions is frequently observed. This could be an indication that, alongside a fundamental consensus, individual perspectives, technical knowledge and understanding, context-specific factors, or personal experiences play a role in the perception of such systems.

## 1 Conflict Potentials Associated with the Use of Applications Based on Artificial Intelligence

Algorithms and methods of Artificial Intelligence (AI) are increasingly employed to personalize information and individualize user experiences (Zhang, Lu & Jin, 2021; Steck, 2021; Batmaz et al., 2019; Mu, 2018). However, in the interaction between humans and AI, conflicts with established design principles can arise (DIN, 2020; Komischke, 2021), partly due to the underlying mathematical probability calculations of intelligent algorithms (Dörn, 2018; Ghahramani, 2015; Sebe et al., 2005) as well as uncertainties in the real world (Russell & Norvig, 2012; Negnevitsky, 2005; Horvitz & Zilberstein, 2001; Sesink, 1993). Furthermore, concerns may arise regarding the evaluation and protection of the data used (Ge et al., 2022; Barenkamp, 2022). A balance between humans and AI-based systems could, however, be achieved by promoting cooperation and communication between both, while simultaneously ensuring that AI-based systems are developed with consideration for user needs (Zhang et al., 2021; Steck, 2011; Batmaz et al., 2019; Mu, 2018; Seufert & Meier, 2023; Amershi et al., 2019). The present study focuses generally on the principles of User-Centered Design (Jokela et al., 2003; Lowdermilk, 2013; Dopp et al, 2019) and, in this context, among others, with usability and, in a broader framework, with the user experience of AI-based interactive systems (Spaulding & Weber, 2009; Bond et al, 2019; Spallazzo, 2022). Specifically, it investigates to what extent specific usability aspects are important for users interacting with such systems, particularly when AI methods are employed.

### 1.1 Thematic Focus

Against this background, the question of "how" is at the center of this study, which focuses in particular on two thematic fields of human-AI collaboration:

[1] *This paper was presented at the GMW Annual Conference 2023 in Jena, Germany.*

(1) Explainability, where the focus lies on the transparency of an AI-based (recommender) system.
(2) Controllability (or steerability), with a focus on the adaptivity of an AI-based (recommender) system.

### 1.2 Background

In a transnational university network of 13 higher education institutions that offer digital degree programs, a dashboard for learner self-regulation is being developed (Drzyzga et al., 2023). Upon the learners' consent, this dashboard is intended to provide personalized evaluations based on their data from the Learning Management System (LMS). The goal of the project is to develop a dashboard that, in addition to providing a possibility for self-regulation, provides learners with individual recommendations and forecasts regarding their learning process based on machine learning methods; to achieve this, data from the LMS are to be merged and analyzed in combination with information on the course of study from the university administration. This provided the rationale for investigating the identified conflicts regarding the usability of AI-based (recommender) systems (see "2 Methodology").

## 2 Method

A questionnaire was chosen to conduct the study. This consisted of two parts: the first part addressed topics of online learning and the support of the learning process through a dashboard for self-regulation with predictive elements based on a data analysis of learning behavior; the second part focused on AI-based (recommender) systems in general. At the time of the survey, there was no concrete dashboard; rather, the goal was to collect initial requirements and solicit the opinions of potential users on this topic. Based on an initial analysis of conventional design principles in the field of Human-Computer Interaction and their application to AI-based (recommender) systems or their user interface design, as well as the question of where discrepancies may occur, identified conflicts were formulated into items for the second part of the survey (Author, in preparation).

For the present study, a total of 12 items from the pilot study were used, which were to be evaluated on a 5-point Likert scale (priority) including an abstention option ("cannot judge") with values from 1 to 5 ("does not apply" (1), "tends not to apply" (2), "partly/partly" (3), "tends to apply" (4), "completely applies" (5)). At the beginning of the questions, participants were given an introductory text on the subject matter. The starting point for this was the consideration of what is generally important to users when operating such systems, without requiring prior experience with the concrete system (see "3.1 Identified Conflicts in Human-AI Collaboration"). The 12 items, distributed equally with 6 items on the two aforementioned thematic fields (cf. Tables 1 and 2 in the following chapter), were presented to students within a university network for online degree programs as part of a ten-day online survey. Prior to the start of the survey, the questionnaire was subjected to both qualitative and quantitative pretesting.

From the tensions identified in the pilot study, a central research question was derived and placed before the items: What is important to users when interacting with (recommender) systems that react individually through the application of certain rules (algorithms) and the use of AI, for example, by analyzing personal data (e.g., demographic data or click behavior) of the users to provide recommendations or generally adjust system behavior?

# 3 Results

A total of 179 valid responses were registered. The demographic data showed a participation consisting of 84 men, 90 women, and 5 diverse individuals. The age groups were distributed as follows: in the age group 18-24 years there were 29 participants, in the age group 25-32 years 62 participants, in the age group 33-50 years 78 participants, and in the age group over 50 years a total of 10 participants. The intended degree is divided into 160 participants pursuing a Bachelor's degree and 19 participants pursuing a Master's degree.

## 3.1 Identified Conflicts in Human-AI Collaboration

From the individual conflicts, thematic aspects with corresponding research questions (items) were derived. Table 1 shows the individual aspects and items for the thematic field “Explainability” and Table 2 shows the individual aspects and items for the thematic field “Controllability”, which were to be answered by the respondents regarding how important the corresponding aspect is to them. The column “Valid” indicates how many participants provided an assessment for the respective aspect.

*Tabelle 1: Aspects, identified conflicts, and values for the assessment of human-AI collaboration in the field of “Explainability”*

| # | Aspect | Item | Valid | Median |
|---|---|---|---|---|
| 1 | Transparent data analysis | ... that I receive explanations as to which data are used for the analysis. | 175 | 4,00 |
| 2 | Functionality of the algorithm used | ... that I receive explanations as to how the algorithm works. | 175 | 4,00 |
| 3 | Data processing | ... that I receive explanations as to how the system evaluates the data. | 175 | 4,00 |
| 4 | Need for explanations of criteria used | ... that I receive explanations as to which criteria led to the displayed result. | 177 | 5,00 |
| 5 | Percentage indication of the applicable result | ... that I receive explanations as to what percentage the result applies to me. | 174 | 4,00 |

| 6 | Context-dependent explanations | ... that I receive context-related explanations. (e.g., in order to improve the result or operate the system.) | 174 | 4,00 |
|---|---|---|---|---|

*Tabelle 2: Aspects, identified conflicts, and values for the assessment of human-AI collaboration in the field of "Controllability"*

| # | Aspect | Item | Valid | Median |
|---|---|---|---|---|
| 1 | Setting of preferences | ... that I can set the system according to my personal interests (pre-configuration). | 175 | 4,00 |
| 2 | Result-based configuration | ... that I can exert immediate influence on the result (subsequently, based on results). | 166 | 4,00 |
| 3 | Adding additional information | ... that I can add additional information to the system (such as information not provided by the system). | 169 | 4,00 |
| 4 | Correspondence with personal self-assessment | ... that the displayed result corresponds at least to my personal self-assessment. | 153 | 3,00 |
| 5 | Dialogue support, automated or with human help | ... that I can ask follow-up questions. (Dialogue support, e.g., use of an automated chat where, if necessary, a human from the providing system can be connected.) | 174 | 4,00 |
| 6 | Restoration of the original state | ... that I can return to the original state after having carried out settings. | 173 | 4,00 |

### 3.2 Thematic Field Explainability

In the following, the thematic field "Explainability" is focused on with six different aspects. The statements of the items point to the importance of background knowledge regarding the functionality of these systems in order to enable users a better understanding and to create a sense of trust.

#### 3.2.1 Transparent Data Analysis

The necessity to bring transparency to AI-based (recommender) systems is demonstrated by providing explanations as to which data are used for an analysis. Like the subsequent item ("Functionality of the algorithm used"), it aims at the necessity of keeping the process of interactions and the processing of personal data within these systems transparent, so that users can understand them better and trust them. It becomes clear that the participants are largely in agreement that this is important (median value of 4), although there are also outliers who consider this unimportant.

### 3.2.2 Functionality of the Algorithm Used

This item reflects an overall slightly positive opinion (median value 4) regarding the functionality of the algorithms used in AI-based (recommender) systems. In summary, it can be stated that most participants consider it important that the functionality of the algorithms used by AI-based (recommender) systems is explained. There are some differences in the assessment of the importance of this topic, but overall there seems to be a moderate agreement among the participants in the rather higher value range of 4.

### 3.2.3 Data Processing

The question of whether AI-based (recommender) systems should be designed such that explanations are given as to how the system evaluates the data is answered tendentially positively by the respondents (median value 4). This concerns traceability, particularly regarding the functionality of the system, i.e., including its algorithms and data processing methods as well as the factors that the system takes into account when creating the result.

### 3.2.4 Need for Explanations of Criteria Used

An overwhelmingly positive opinion (median value 5) is shown regarding the question of whether explanations are given regarding the criteria that led to the displayed result.

### 3.2.5 Percentage Indication of the Applicable Result

Overall, a slightly positive opinion (median value 4) results for this item, which is intended to provide information as to how important it is to users to be explained to what percentage the result of an AI-based (recommender) system applies to them. While most respondents consider this rather important, there is also a certain neutral spectrum of opinion regarding this question.

### 3.2.6 Context-Dependent Explanations

The opinion regarding the need for context-related explanations for improving or effectively using the system is overall slightly positive (median value 4). This question provides insight into whether users desire more information and clarity regarding how these systems function and how they can benefit from them. Overall, the results of this item indicate that most opinions consider context-related explanations important in order to improve or better use AI-based (recommender) systems. However, there is a large range of opinions regarding this question (lower whisker at value 1), i.e., some respondents do not consider this important at all.

## 3.3 Thematic Field Controllability

In the following, the thematic field „Controllability" is focused on with six different aspects. The statements of the items aim at the necessity of designing the functionality of these systems in an expectation-conforming manner - traceable - so that users can understand them better and trust them.

### 3.3.1 Setting of Preferences

Pre-configuration, i.e., users should be able to specify their preferences in advance so that the system can generate correspondingly personalized results. The goal of this approach is to give users a certain degree of control over the use and personalization of the AI-based (recommender) system. Overall, the results show a rather positive view (median value 4) of personalized pre-configuration for such systems, but also a certain variance in preferences and opinions regarding importance.

### 3.3.2 Result-Based Configuration

An overall slightly positive opinion (median value 4) exists regarding retroactive influence or result-based configuration, where users should be given the possibility to change or modify a result after it has been generated by the system based on their feedback. This statement refers to subsequent adjustments and fine-tuning of the results of AI-based (recommender) systems. In summary, it can be said that while most participants value having direct influence on the results, this is not necessarily a (top) priority for everyone.

### 3.3.3 Adding Additional Information

There is an overall slightly positive opinion (median value 4) that users want to be able to add their own information or preferences in order to improve the accuracy of the result. However, there are also more neutral opinions.

### 3.3.4 Correspondence with Personal Self-Assessment

The item aims at the importance of personal self-assessment in determining the relevance of the results displayed by the system. With a median value of 3, the assessment lies in the rather neutral range. Overall, the results indicate a generally neutral view of personal self-assessment when evaluating the displayed results, but also some differences in opinions and preferences regarding importance - notably, there is a high number of abstentions (n=26) here.

### 3.3.5 Dialogue Support, Automated or with Human Help

The item aims at the preference for flexibility and control of the AI-based (recommender) system. It particularly underscores the need of users for autonomy, specifically autonomy within the system. With a median value of 4, the assessment is overall in the rather slightly positive range. Regarding human interaction in these systems, users seem to appreciate a certain degree of dialogue support. This can occur either automatically or through human assistance.

### 3.3.6 Restoration of the Original State

As with the previous item, this also aims at a preference for flexibility and control of the AI-based (recommender) system. This statement refers to the possibility of being able to make adjustments (control) in order to return to the original state. With a median value of 4, the assessment is overall in the rather slightly positive range. In general, it seems to be important for the majority of respondents that AI-based (recommender) systems offer the possibility to return to the original settings.

# 4 Discussion

The results show that the aspects with the identified conflicts were generally classified as important by the respondents (median values: 1 x 5.0, 10 x 4.00, 1 x 3.00). This shows an overall clear tendency that the thematic fields of explainability and controllability contain points that are classified as rather important. However, a high dispersion of ratings is frequently observed. This high dispersion, despite predominantly uniform response behavior of the respondents, could be attributed to different conceptions regarding how these aspects should be achieved or implemented in an AI-based (recommender) system: For example, there could be a significant discrepancy between the understanding and the expectations of users for an understandable AI-based (recommender) system, which could lead to different interpretations and opinions regarding the importance or the benefit regarding the use of such systems. Furthermore, individual perspectives (what constitutes an "explainable" system or how much transparency is required for an AI-based (recommender) system), technical knowledge and understanding (degree of experience in interacting with such systems), context-specific factors (the importance of explainability can vary depending on application or industry), or personal experiences could play a role. For example, positive or negative experiences with these systems could have influenced the expectations and preferences towards AI-based (recommender) systems.

The widespread agreement of the respondents in both thematic fields, however, again points to a common conception or a common knowledge about the importance and the goals of understandable AI-based (recommender) systems. This could be an indication that the respondents are aware of the potential risks that can be associated with opaque AI models, such as lack of trust in their decisions or ethical concerns during interaction with such systems.

In conclusion, it can be stated that these results could also reflect the prior knowledge of the participants, which must also be taken into account in view of the limitations of this study due to the sample size. The strength of the study lies in its focused approach, which is based on discrepancies in the interaction between humans and AI-based (recommender) systems. The restriction to students proved positive for the goal of investigating the interaction with AI-based (recommender) systems in the higher education sector, but it could also be limiting regarding the generalizability of the results, so that a further investigation is planned. The results will continue to serve as a basis for future investigations.

*German version*

# Mensch und Künstliche Intelligenz - eine vertrauensvolle und verständliche Zusammenarbeit fördern

*Gilbert Drzyzga*

Methoden der Künstlichen Intelligenz (KI) ermöglichen in vielen Bereichen die Personalisierung von Informationen für individuelle Nutzungserfahrungen, können aber auch in Konflikt mit bestehenden Gestaltungsprinzipien geraten, z. B. aufgrund von Unsicherheiten in Bezug auf die reale Welt. Der Aufbau von Vertrauen und Verständnis kann ein Ansatz sein, um ein ausgewogeneres Verhältnis zwischen Mensch und KI zu schaffen. Aufbauend auf einer Vorstudie wurden in einer Online-Befragung 12 Einzelaspekte zu den Themen Erklärbarkeit und Steuerbarkeit untersucht. Die Ergebnisse zeigen, dass beide Themen mit ihren unterschiedlichen und zahlreichen Facetten von den Befragten tendenziell als wichtig erachtet werden, gleichzeitig aber häufig eine breite Streuung der Meinungen zu beobachten ist. Dies könnte ein Hinweis darauf sein, dass neben einem grundsätzlichen Konsens auch individuelle Sichtweisen, technisches Wissen und Verständnis, kontextspezifische Faktoren oder persönliche Erfahrungen bei der Wahrnehmung solcher Systeme eine Rolle spielen.

## 1 Konfliktpotenziale im Zusammenhang mit dem Einsatz von Anwendungen, die auf Künstlicher Intelligenz basieren

Algorithmen und Methoden der Künstlichen Intelligenz (KI) werden zunehmend eingesetzt, um Informationen zu personalisieren und die Nutzungserfahrungen zu individualisieren (Zhang, Lu & Jin, 2021; Steck, 2021; Batmaz et al., 2019; Mu, 2018). Allerdings kann es in der Interaktion zwischen Mensch und KI unter anderem infolge der den intelligenten Algorithmen zugrundeliegenden mathematischen Wahrscheinlichkeitsberechnungen (Dörn, 2018; Ghahramani, 2015; Sebe et. al, 2005) sowie Unsicherheiten in der realen Welt (Russell & Norvig, 2012; Negnevitsky, 2005; Horvitz & Zilberstein, 2001; Sesink, 1993) zu Konflikten mit bestehenden Gestaltungsprinzipien kommen (DIN, 2020; Komischke, 2021). Weiterhin können sich Bedenken hinsichtlich der Bewertung und des Schutzes der verwendeten Daten ergeben (Ge et al., 2022; Barenkamp, 2022). Ein Gleichgewicht zwischen Menschen und KI-basierten Systemen könnte jedoch erreicht werden, indem die Zusammenarbeit und Kommunikation zwischen beiden gefördert und gleichzeitig sichergestellt wird, dass KI-basierte Systeme unter Berücksichtigung der Bedürfnisse der Nutzenden entwickelt werden (Zhang et al., 2021; Steck, 2011; Batmaz et al., 2019; Mu, 2018; Seufert & Meier, 2023; Amershi et al., 2019). Die vorliegende Studie beschäftigt sich allgemein mit den Prinzipien des User-Centered Design (Jokela et al., 2003; Lowdermilk, 2013; Dopp et al, 2019) und damit in diesem Kontext u. a. mit der Usability und im weiteren Rahmen mit der User Experience von KI-basierten interaktiven Systemen (Spaulding & Weber, 2009; Bond et al, 2019; Spallazzo, 2022), d. h. es wird untersucht, inwiefern spezifische Aspekte der Gebrauchstauglichkeit für die Nutzenden bei der Verwendung solcher Systeme, insbesondere bei Einsatz von KI-Methoden, wichtig sind.

### 1.1 Thematischer Fokus

Vor diesem Hintergrund steht die Frage nach dem „Wie“ im Mittelpunkt dieser Studie, die sich dazu insbesondere auf zwei Themenfelder der Zusammenarbeit von Mensch und KI konzentriert:

(1) Erklärbarkeit, wobei die Fokussierung auf der Transparenz eines KI-basierten (Empfehlungs-)Systems liegt.
(2) Kontrollierbarkeit bzw. Steuerbarkeit mit Fokus auf die Adaptivität eines KI-basierten (Empfehlungs-)Systems.

### 1.2 Hintergrund

In einem Hochschulnetzwerk von 13 länderübergreifenden Hochschulen, die digitale Studiengänge anbieten, wird ein Dashboard für Lernende zur Selbstregulation entwickelt (Drzyzga et al., 2023). Dieses Dashboard soll den Lernenden nach deren Zustimmung u.a. personalisierte Auswertungen auf Basis ihrer Daten aus dem Lernmanagementsystem (LMS) zur Verfügung stellen. Das Ziel des Projektes ist es, ein Dashboard zu entwickeln, das neben einer Möglichkeit zur Selbstregulation, den Lernenden u.a. individuelle Empfehlungen und Prognosen zu ihrem Lernprozess auf Basis maschineller Lernverfahren bereitstellt, dazu sollen die Daten aus dem LMS in Kombination mit den Informationen zum Studienverlauf aus der Hochschulverwaltung zusammengeführt und analysiert werden. Dies wurde als Anlass genommen, die identifizierten Konflikte in Bezug auf die Gebrauchstauglichkeit von KI-basierten (Empfehlungs-)Systemen zu untersuchen (siehe hierzu „2 Methodik“).

## 2 Methodik

Zur Durchführung der Studie wurde die Form eines Fragebogens gewählt. Dieser bestand aus zwei Teilen: Der erste Teil befasste sich mit Themen des Online-Lernens und der Unterstützung des Lernprozesses durch ein Dashboard zur Selbstregulation mit prädiktiven Elementen auf Basis einer Datenanalyse des Lernverhaltens, der zweite Teil konzentrierte sich auf KI-basierte (Empfehlungs-)Systeme im Allgemeinen. Zum Zeitpunkt der Befragung gab es noch kein konkretes Dashboard, vielmehr ging es darum, erste Anforderungen zu sammeln und die Meinung der potenziellen Nutzenden zu diesem Thema einzuholen. Ausgehend von einer ersten Analyse konventioneller Gestaltungsprinzipien im Bereich der Mensch-Computer-Interaktion und deren Anwendung auf KI-basierte (Empfehlungs-)Systeme bzw. deren User-Interface-Gestaltung und der Frage, wo dabei Diskrepanzen auftreten können, wurden identifizierte Konflikte für den zweiten Teil der Befragung in Form von Items formuliert (Verfasser, in Vorbereitung).

Für die vorliegende Studie wurden insgesamt 12 Items aus der Vorstudie verwendet, die auf einer 5-Punkte-Likert-Skala (Priorität) inkl. Enthaltungsoption („kann ich nicht beurteilen“) mit den Werten 1 bis 5 („Trifft nicht zu“ (1), „Trifft eher nicht zu“ (2), „teils/teils“ (3), „Trifft eher zu“ (4), „Trifft vollkommen zu“ (5)) zu bewerten waren. Zu Beginn der Fragen wurde den Teilnehmenden ein einführender Text zur Thematik gegeben. Ausgangspunkt dazu war die Überlegung, was den Nutzenden bei der Bedienung solcher Systeme generell wichtig ist, ohne dass Erfahrungen mit dem konkreten System vorliegen mussten (siehe dazu „3.1 Identifizierte Konflikte der Mensch-KI-Kollaboration“). Die 12 Items, die sich mit jeweils 6 Items auf die beiden genannten Themenfelder verteilen (vgl. dazu die Tabellen 1 und 2 im folgenden Kapitel), wurden den Studierenden im Rahmen einer zehntägigen Online-Befragung innerhalb eines Hochschulverbundes für Online-Studiengänge vorgelegt. Vor Beginn der Befragung wurde der Fragebogen sowohl einem qualitativen als auch einem quantitativen Pretest unterzogen.
Aus den in der Vorstudie identifizierten Spannungsfeldern wurde eine zentrale Fragestellung abgeleitet und den Items vorangestellt: Was ist für die Nutzenden wichtig im Umgang mit (Empfehlungs-)Systemen, die durch die Anwendung bestimmter Regeln (Algorithmen) und den Einsatz von KI, z. B. durch die Analyse persönlicher Daten (z. B. demographische Daten oder Klickverhalten) der Nutzenden, individualisiert reagieren (z. B. Empfehlungen zu etwas geben oder das Systemverhalten generell anpassen).

# 3 Ergebnisse

Insgesamt wurden 179 gültige Antworten registriert. Die Angaben zu den demographischen Daten ergaben eine Teilnahme von insgesamt 84 Männern, 90 Frauen und 5 Diversen. Die Altersgruppen verteilten sich wie folgt: In der Altersgruppe 18-24 Jahre waren es 29 Teilnehmende, in der Altersgruppe 25-32 Jahre 62 Teilnehmende, in der Altersgruppe 33-50 Jahre 78 Teilnehmende und in der Altersgruppe über 50 Jahre insgesamt 10 Teilnehmende. Der angestrebte Abschluss teilt sich auf in 160 Teilnehmende, die einen Bachelor-Abschluss anstreben und 19 Teilnehmende, die einen Master-Abschluss anstreben.

## 3.1 Identifizierte Konflikte der Mensch-KI-Kollaboration

Aus den einzelnen Konflikten wurden Themenaspekte mit entsprechenden Fragestellungen (Items) abgeleitet. Die Tabelle 1 zeigt die einzelnen Aspekte und Items zum Themenfeld „Erklärbarkeit“ und die Tabelle 2 die einzelnen Aspekte und Items zum Themenfeld „Steuerbarkeit“, die von den Befragten jeweils dahingehend beantwortet werden sollten, wie wichtig der entsprechende Aspekt für sie ist. Die Spalte „Gültig“ gibt an, wie viele Teilnehmende für den jeweiligen Aspekt eine Bewertung abgegeben haben.

*Tabelle 1: Aspekte, identifizierte Konflikte und Werte zur Einschätzung der Mensch-KI-Zusammenarbeit im Themenfeld „Erklärbarkeit“*

| # | Aspekt | Item | Gültig | Median |
|---|---|---|---|---|
| 1 | Transparente Datenanalyse | ... dass ich Erklärungen dazu erhalte, welche Daten für die Analyse verwendet werden. | 175 | 4,00 |
| 2 | Funktionsweise des eingesetzten Algorithmus | ... dass ich Erklärungen dazu erhalte, wie der Algorithmus arbeitet. | 175 | 4,00 |
| 3 | Datenverarbeitung | ... dass ich Erklärungen dazu erhalte, wie das System die Daten auswertet. | 175 | 4,00 |
| 4 | Bedarf an Erklärungen zu den verwendeten Kriterien | ... dass ich Erklärungen dazu erhalte, welche Kriterien zum angezeigten Ergebnis geführt haben. | 177 | 5,00 |
| 5 | Prozentuale Angabe des zutreffenden Ergebnisses | ... dass ich Erklärungen dazu erhalte, zu wie viel Prozent das Ergebnis auf mich zutrifft. | 174 | 4,00 |
| 6 | Kontextabhängige Erklärungen | ... dass ich kontextbezogene Erklärungen erhalte. (Um z. B. das Ergebnis zu verbessern oder das System zu bedienen.) | 174 | 4,00 |

*Tabelle 2: Aspekte, identifizierte Konflikte und Werte zur Einschätzung der Mensch-KI-Zusammenarbeit im Themenfeld „Steuerbarkeit"*

| # | Aspekt | Item | Gültig | Median |
|---|---|---|---|---|
| 1 | Festlegung von Präferenzen | ... dass ich das System nach meinen persönlichen Interessen einstellen kann (Vorkonfiguration). | 175 | 4,00 |
| 2 | Ergebnisbasierte Konfiguration | ... dass ich unmittelbar Einfluss auf das Ergebnis nehmen kann (nachträglich, ergebnisbezogen). | 166 | 4,00 |
| 3 | Hinzufügen zusätzlicher Informationen | ... dass ich zusätzliche Informationen zum System hinzufügen kann (solche, die nicht vom System vorgesehen waren). | 169 | 4,00 |
| 4 | Entsprechung der persönlichen Selbsteinschätzung | ... dass das dargestellte Ergebnis mindestens meiner persönlichen Selbsteinschätzung entspricht. | 153 | 3,00 |
| 5 | Dialogunterstützung, automatisiert oder mit menschlicher Hilfe | ... dass ich Rückfragen stellen kann. (Dialogunterstützung, z. B. Nutzung eines automatisierten Chats, bei dem bei Bedarf ein Mensch | 174 | 4,00 |
| | | des anbietenden Systems zugeschaltet werden kann.) | | |
| 6 | Wiederherstellung des Originalzustandes | ... dass ich nach durchgeführten Einstellungen wieder in den Originalzustand zurückkehren kann. | 173 | 4,00 |

## 3.2 Themenfeld Erklärbarkeit

Im Folgenden wird das Themenfeld „Erklärbarkeit“ mit sechs verschiedenen Aspekten fokussiert. Die Aussagen der Items weisen auf die Bedeutung des Hintergrundwissens zur Funktionsweise dieser Systeme hin, um den Nutzenden ein besseres Verständnis zu ermöglichen und ein Gefühl des Vertrauens zu schaffen.

### 3.2.1 Transparente Datenanalyse

Die Notwendigkeit, Transparenz in KI-basierte (Empfehlungs-)Systeme zu bringen, zeigt sich dadurch, Erklärungen zur Verfügung zu stellen, welche Daten für eine Analyse verwendet werden. Wie auch das nachfolgende Item („Funktionsweise des eingesetzten Algorithmus“) zielt es auf die Notwendigkeit ab, den Ablauf der Interaktionen und die Verarbeitung persönlicher Daten innerhalb dieser Systeme transparent zu halten, damit die Nutzenden sie besser verstehen und ihnen vertrauen können. Es wird deutlich, dass sich die Teilnehmenden weitgehend einig sind, dass dies wichtig ist (Medianwert von 4), dass es aber auch Ausreißer gibt, die dies als unwichtig erachten.

### 3.2.2 Funktionsweise des eingesetzten Algorithmus

Dieses Item gibt ein insgesamt leicht positives Meinungsbild (Medianwert 4) bezüglich der Funktionsweise der in KI-basierten (Empfehlungs-)Systemen verwendeten Algorithmen wieder. Zusammenfassend lässt sich feststellen, dass die meisten Teilnehmenden es für wichtig halten, dass die Funktionsweise der verwendeten Algorithmen von KI-basierten (Empfehlungs-)Systemen erklärt wird. Es gibt einige Unterschiede in der Einschätzung der Wichtigkeit dieses Themas, aber insgesamt scheint es eine moderate Übereinstimmung unter den Teilnehmenden im eher höheren Wertebereich von 4 zu geben.

### 3.2.3 Datenverarbeitung

Die Frage, ob KI-basierte (Empfehlungs-)Systeme so gestaltet sein sollten, dass Erklärungen dazu gegeben werden, wie das System die Daten auswertet, wird von den Befragten tendenziell positiv beantwortet (Medianwert 4). Dabei geht es um die Nachvollziehbarkeit insbesondere hinsichtlich der Funktionsweise des Systems, d. h. einschließlich seiner Algorithmen und Datenverarbeitungsmethoden sowie der Faktoren, die das System bei der Erstellung des Ergebnisses berücksichtigt.

### 3.2.4 Bedarf an Erklärungen zu den verwendeten Kriterien

Ein überwiegend positives Meinungsbild (Medianwert 5) zeigt sich bei der Frage, ob Erläuterungen zu den Kriterien gegeben werden, die zu dem angezeigten Ergebnis geführt haben.

### 3.2.5 Prozentuale Angabe des zutreffenden Ergebnisses

Insgesamt ergibt sich ein leicht positives Meinungsbild (Medianwert 4) zu diesem Item, das Auskunft darüber geben soll, wie wichtig es den Nutzenden ist, erklärt zu bekommen, zu wie viel Prozent das Ergebnis eines KI-basierten (Empfehlungs-)Systems auf sie zutrifft. Während die meisten Befragten dies als eher wichtig erachten, gibt es auch ein gewisses neutrales Meinungsspektrum zu dieser Frage.

### 3.2.6 Kontextabhängige Erklärungen

Die Meinung über den Bedarf an kontextbezogenen Erklärungen zur Verbesserung oder effektiven Nutzung des Systems ist insgesamt leicht positiv (Medianwert 4). Diese Frage gibt insgesamt Aufschluss darüber, ob die Nutzenden mehr Informationen und Klarheit darüber wünschen, wie diese Systeme funktionieren und wie sie davon profitieren können. Insgesamt deuten die Ergebnisse dieses Items darauf hin, dass die meisten Meinungen kontextbezogene Erklärungen für wichtig halten, um KI-basierte (Empfehlungs-)Systeme zu verbessern oder besser nutzen zu können. Allerdings gibt es bei dieser Frage eine große Bandbreite an Meinungen (unterer Whisker bei Wert 1), d. h. einige Befragte halten dies überhaupt nicht für wichtig.

## 3.3 Themenfeld Steuerbarkeit

Im Folgenden wird der Themenfeld „Steuerbarkeit“ mit sechs verschiedenen Aspekten fokussiert. Die Aussagen der Items zielen auf die Notwendigkeit, die Funktionsweise dieser Systeme erwartungskonform - nachvollziehbar - zu gestalten, damit die Nutzenden sie besser verstehen und ihnen vertrauen können.

### 3.3.1 Festlegung von Präferenzen

Vorkonfiguration, d. h. die Nutzenden sollten in der Lage sein, ihre Präferenzen im Voraus anzugeben, damit das System entsprechend personalisierte Ergebnisse generieren kann. Ziel dieses Ansatzes ist es, den Nutzenden ein gewisses Maß an Kontrolle über die Nutzung und Personalisierung des KI-basierten (Empfehlungs-)Systems zu geben. Insgesamt zeigen die Ergebnisse eine eher positive Sichtweise (Medianwert 4) der personalisierten Vorkonfiguration für solche Systeme, aber auch eine gewisse Varianz der Präferenzen und Meinungen hinsichtlich der Wichtigkeit.

### 3.3.2 Ergebnisbasierte Konfiguration

Ein insgesamt leicht positives Meinungsbild (Medianwert 4) gibt es in Bezug auf die rückwirkende Einflussnahme bzw. ergebnisbasierte Konfiguration, bei der die Nutzenden die Möglichkeit gegeben werden soll, ein Ergebnis, nachdem es vom System auf Basis ihres Feedbacks generiert wurde, zu verändern oder zu modifizieren. Diese Aussage bezieht sich auf nachträgliche Anpassungen und Feinabstimmungen der Ergebnisse von KI-basierten (Empfehlungs-)Systemen. Zusammenfassend lässt sich sagen, dass die meisten Teilnehmenden zwar Wert darauf legen, direkten Einfluss auf die Ergebnisse zu haben, dies aber nicht unbedingt für alle eine (oberste) Priorität darstellt.

### 3.3.3 Hinzufügen zusätzlicher Informationen

Es gibt eine insgesamt leicht positive Meinung (Medianwert 4) darüber, dass die Nutzenden in der Lage sein wollen, ihre eigenen Informationen oder Präferenzen hinzuzufügen, um die Genauigkeit des Ergebnisses zu verbessern. Es gibt jedoch auch neutralere Meinungen.

### 3.3.4 Entsprechung der persönlichen Selbsteinschätzung

Das Item zielt auf die Bedeutung der persönlichen Selbsteinschätzung bei der Bestimmung der Relevanz der vom System angezeigten Ergebnisse ab. Mit einem Medianwert von 3 liegt die Bewertung dieses Items im eher neutralen Bereich. Insgesamt deuten die Ergebnisse auf eine generell neutrale Sichtweise der persönlichen Selbsteinschätzung bei der Bewertung der angezeigten Ergebnisse hin, aber auch auf einige Unterschiede in den Meinungen und Präferenzen hinsichtlich der Wichtigkeit - auffallend ist hier die hohe Anzahl an Enthaltungen (n=26).

### 3.3.5 Dialogunterstützung, automatisiert oder mit menschlicher Hilfe

Das Item zielt auf die Präferenz für Flexibilität und Kontrolle des KI-basierten (Empfehlungs-)Systems ab. Es unterstreicht insbesondere das Bedürfnis der Nutzenden nach Autonomie bzw. Autonomie innerhalb des Systems. Mit einem Medianwert von 4 liegt die Bewertung dieses Items insgesamt im eher leicht positiven Bereich. Was die menschliche Interaktion in diesen Systemen betrifft, so scheinen die Nutzenden ein gewisses Maß an Dialogunterstützung zu schätzen. Diese kann entweder automatisch oder durch menschliche Unterstützung erfolgen.

### 3.3.6 Wiederherstellung des Originalzustandes

Wie das vorhergehende Item zielt auch dieses auf eine Präferenz für Flexibilität und Kontrolle des KI-basierten (Empfehlungs-)Systems ab. Diese Aussage bezieht sich auf die Möglichkeit, Anpassungen (Kontrolle) vornehmen zu können, um zum Ausgangszustand zurückzukehren. Mit einem Medianwert von 4 liegt die Bewertung dieses Items insgesamt im eher leicht positiven Bereich. Generell scheint es für die Mehrheit der Befragten wichtig zu sein, dass KI-basierte (Empfehlungs-)Systeme die Möglichkeit bieten, zu den ursprünglichen Einstellungen zurückzukehren.

## 4 Diskussion

Die Ergebnisse zeigen, dass die Aspekte mit den identifizierten Konflikten von den Befragten im Allgemeinen als wichtig eingestuft wurden (Medianwerte: 1 x 5,0, 10 x 4,00, 1 x 3,00). Damit zeigt sich insgesamt eine deutliche Tendenz, dass die Themenfelder Erklärbarkeit und Steuerbarkeit mit ihren unterschiedlichen Ausprägungen durchaus Punkte enthalten, die als eher wichtig eingestuft wurden.

Allerdings ist häufig auch eine hohe Streuung der Bewertungen zu beobachten. Diese hohe Streuung bei überwiegend einheitlichem Antwortverhalten der Befragten könnte auf unterschiedliche Vorstellungen darüber zurückzuführen sein, wie diese Aspekte in einem KI-basierten (Empfehlungs-)System erreicht bzw. umgesetzt werden sollten: So könnte es z. B. eine signifikante Diskrepanz zwischen dem Verständnis und den Erwartungen der Nutzenden an ein verständliches KI-basiertes (Empfehlungs-)System geben, was zu unterschiedlichen Interpretationen und Meinungen über die Bedeutung bzw. den Nutzen hinsichtlich der Nutzung solcher Systeme führen könnte. Darüber hinaus könnten individuelle Sichtweisen (was macht ein „erklärbares" System aus oder wie viel Transparenz ist für ein KI-basiertes (Empfehlungs-)System erforderlich), technisches Wissen und Verständnis (Grad der Erfahrung im Umgang mit solchen Systemen), kontextspezifische Faktoren (die Bedeutung der Erklärbarkeit kann je nach Anwendung oder Branche variieren) oder persönliche Erfahrungen eine Rolle spielen. So könnten beispielsweise positive oder negative Erfahrungen mit diesen Systemen die Erwartungen und Präferenzen gegenüber KI-basierten (Empfehlungs-)Systemen beeinflusst haben.

Die weitgehende Übereinstimmung der Befragten in beiden Themenfeldern deutet jedoch wiederum auf eine gemeinsame Vorstellung bzw. ein gemeinsames Wissen über die Bedeutung und die Ziele verständlicher KI-basierter (Empfehlungs-)Systeme hin. Dies kann ein Hinweis darauf sein, dass sich die Befragten der potenziellen Risiken bewusst sind, die mit undurchsichtigen KI-Modellen verbunden sein können, wie z. B. mangelndes Vertrauen in ihre Entscheidungen oder ethische Bedenken bei der Interaktion mit solchen Systemen.

Abschließend kann festgehalten werden, dass diese Ergebnisse u.a. auch das Vorwissen der Teilnehmenden widerspiegeln könnten, was auch im Hinblick auf die Grenzen dieser Studie aufgrund der Stichprobengröße berücksichtigt werden muss. Die Stärke der Studie liegt in ihrem fokussierten Ansatz, der auf Diskrepanzen in der Interaktion zwischen Mensch und KI-basierten (Empfehlungs-)System basiert. Die Beschränkung auf Studierende erwies sich für das Ziel, die Interaktion mit KI-basierten (Empfehlungs-)Systemen im Hochschulbereich zu untersuchen, als positiv, könnte sich aber auch einschränkend auf die Verallgemeinerbarkeit der Ergebnisse auswirken, so dass eine weitere Untersuchung geplant ist. Die Ergebnisse werden weiterhin als Grundlage für zukünftige Untersuchungen dienen.

## Literatur